\documentclass[twocolumn,epsf,graphics,psfig,superscriptaddress]{revtex4}

\usepackage{graphicx,graphics} 
\usepackage[dvips,bookmarks=false]{hyperref}
\usepackage{graphicx}
\usepackage{dcolumn}
\usepackage{eucal}
\usepackage[dvips]{epsfig}

\begin{document}
\title{Modeling of gas adsorption on graphene nanoribbons}
\author{Alireza Saffarzadeh} \email{a-saffar@tpnu.ac.ir}
\affiliation{Department of Physics, Payame Noor University,
Nejatollahi Street, 159995-7613 Tehran, Iran}
\affiliation{Computational Physical Sciences Laboratory,
Department of Nano-Science, Institute for Research in Fundamental
Sciences (IPM), P.O. Box 19395-5531, Tehran, Iran}
\date{\today}

\begin{abstract}
We present a theory to study gas molecules adsorption on armchair
graphene nanoribbons (AGNRs) by applying the results of \emph{ab}
\emph{initio} calculations to the single-band tight-binding
approximation. In addition, the effect of edge states on the
electronic properties of AGNR is included in the calculations.
Under the assumption that the gas molecules adsorb on the ribbon
sites with uniform probability distribution, the applicability of
the method is examined for finite concentrations of adsorption of
several simple gas molecules (CO, NO, CO$_2$, NH$_3$) on 10-AGNR.
We show that the states contributed by the adsorbed CO and NO
molecules are quite localized near the center of original band gap
and suggest that the charge transport in such systems cannot be
enhanced considerably, while CO$_2$ and NH$_3$ molecules
adsorption acts as acceptor and donor, respectively. The results
of this theory at low gas concentration are in good agreement with
those obtained by density-functional theory calculations.
\end{abstract}
\maketitle

\section{Introduction}
In the past two decades carbon nanostructures, such as carbon
nanotubes (CNTs) and graphene, have attracted much attention due
to their novel fundamental properties and possible applications in
future carbon-based nanoelectronics \cite{Saito}. An important
feature of these materials is their sensing property and on this
basis, gas sensors \cite{Watson,Capone}, pH sensors \cite{Ang},
and biosensors \cite{Wu} have been proposed. Among various types
of sensors, most of the experimental
\cite{Kong,Collins,Li,Chopra,Schedin,Romero,Wehling,Lu} and
theoretical
\cite{Zhao,Santucci,Robinson,Peng1,Peng2,Cao,Moradian1,Leenaerts,Zhang}
studies have been focused on the gas sensors characteristics of
the carbon nanostructures due to their promising applications in
various fields such as electronics, agriculture and medicine
\cite{Watson,Capone}.

It has been demonstrated that the CNTs can be used as chemical
sensors for detecting very small concentration of NO$_2$, NH$_3$
and other gases with high sensitivity at room temperature
\cite{Kong,Collins,Li,Chopra}. The interaction of graphene with
chemical environments and the possibility of detecting individual
molecules have also been reported \cite{Schedin,Romero,Wehling}.
The results indicate that the sensing mechanism is based on
changes in the charge carrier concentration induced by gas
molecules adsorbed on the graphene surface and acting as donors or
acceptors. For instance, Leenaerts \textit{et al}
\cite{Leenaerts}, based on the first-principle calculations,
showed that H$_2$O and NO$_2$ behave as acceptors while NH$_3$, CO
and NO act as donors. They found that, molecular doping, i.e.
charge transfer between the adsorbates and the graphene surface,
is almost independent of the adsorption site, which can be
attributed to the translational symmetry of the system. In fact,
due to the flat structure and thus larger accessible surface area,
graphene as well as graphene nanoribbons (GNRs) may act better
than any other carbon-based materials as gas sensors.

The special electronic behavior of GNRs corresponds to the two
typical topological shapes of the carbon atoms on their edges,
namely armchair and zigzag. These ribbons can be either
semiconducting with a size dependent gap or metallic \cite{Han1}.
The first-principles calculations and the results of simple
tight-binding approximation show that all AGNRs with edge
deformation are semiconducting with a finite band gap
\cite{Son1,Zheng}. Due to the semiconducting feature of the AGNRs,
their sensing properties were investigated recently, based on the
tight-binding approximation \cite{Rosales,Moradian2} and the
\textit{ab initio} calculations \cite{Huang}. By changing the
values of hopping integrals and atomic on-site energies for
simulating the gas adsorption on the edges of an AGNR and using
the coherent potential approximation for studying the effect of
finite concentration of gas molecules, the sensing properties of
the AGNR were discussed in Ref. \cite{Moradian2}. Based on density
functional theory (DFT) calculations, Huang and co-workers
\cite{Huang} studied the adsorption of CO, NO, NO$_2$, O$_2$,
N$_2$, CO$_2$, and NH$_3$ on the edges of an AGNR in the presence
of dangling bond defects. The gas adsorption only around the
defect sites, centered at the edges, was considered. They found
that the adsorption of CO$_2$ and O$_2$ molecules changes the AGNR
to $p$-type semiconductors, while NH$_3$ adsorption changes the
system to an $n$-type semiconductor.

We should note that although the \textit{ab initio} techniques
based on DFT obtain a good degree of accuracy to describe the
electronic properties of pristine materials, it is somewhat
restricted when we consider disordered systems such as gas
sensors. The reason is that, in a real disorder system such as gas
adsorption on GNR, one should consider all of the configurations
and obtain a physical quantity by averaging over all possible
adsorption configurations, while in the DFT calculations only one
configuration is usually considered.

In order to fully understand the effects of adsorbed molecules on
the electronic properties of AGNRs, we should consider all
possible configurations of gas adsorption, i.e. the variation of
the position of the gas molecule across the ribbon width.
Therefore, using the values of bond lengths between the AGNR and
the molecules, and also the values of inter-atomic distances
obtained from the DFT calculations \cite{Huang} and using a
scaling rule of the tight-binding hopping integrals, we study the
effects of CO, NO, CO$_2$ and NH$_3$ adsorption on the average
local density of states (LDOS) at different gas concentrations
within the single-band tight-binding approximation. The paper is
organized as follows. In Sec. II, we present the model and
formalism to obtain the Green's function and hence the average
LDOS of the system. The total Green's function is obtained by
means of the iterative procedure \cite{Lopez}, which provides a
quick way of evaluating the surface and bulk Green's functions.
Numerical results and discussions for the electronic properties of
AGNR with gas adsorption are presented in Sec. III. We conclude
our findings in Sec. IV.

\begin{figure}
\centerline{\includegraphics[width=0.9\linewidth]{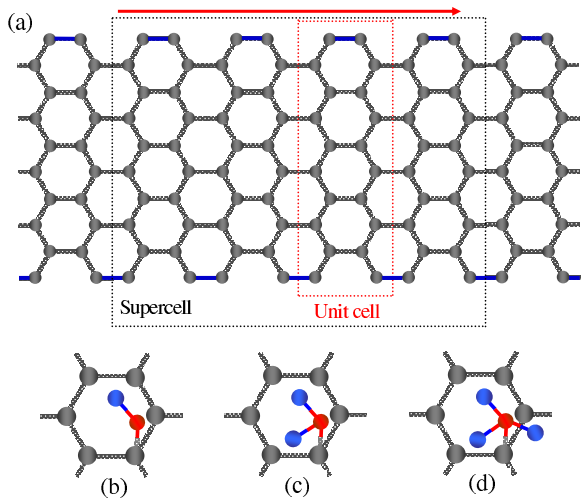}}
\caption{(Color online) (a) Schematic view of the pristine 10-AGNR
with modification in bond lengths at the edges (blue lines). The
arrow shows the periodic direction and the red (black) dotted
rectangle indicates an unit cell (a supercell). (b)-(d) The
structure around adsorbed molecules after the adsorption of
diatomic, triatomic and quadratomic gas molecules, respectively.
Only a single atom (red atom) of each molecule is connected to the
ribbon.}
\end{figure}

\section{Model and formalism}
We consider an AGNR with gas molecules randomly adsorbed on its
surface. The system is described as an infinite stack of
supercells with nearest-neighbor interaction. This means that we
transform the original system into a linear chain of supercells.
Each supercell may contain several unit cells and the gas
molecules adsorb to the AGNR sites with uniform probability. Under
the assumption that only a single molecule adsorbs to each
supercell, the number of supercell carbon atoms, $N_s$, determines
the gas molecule concentration as $x=1/N_s$. The ribbon consists
of $N_a=10$ dimer lines across the ribbon width, as shown in Fig.
1(a), and following conventional notation \cite{Nakada}, the
ribbon is referred as 10-AGNR. We assume that the dangling
$\sigma$ bonds at the edges have been passivated by hydrogen atoms
(not shown), hence we expect the bond lengths between carbon atoms
at the edges differ from those in the interior of the ribbon (see
the blue bonds in Fig. 1(a)). In addition, the gas molecules can
be diatomic, triatomic or quadratomic and we assume that each gas
molecule adsorbs to only one carbon atom (i.e., single contact) as
shown in Fig. 1(b)-1(d). In such a device, the electronic
structure of entire system can be described by a single-electron
Hamiltonian in a basis of localized atomic orbitals for the atoms
in the molecule and $\pi$-orbitals in the 10-AGNR. We use the
tight-binding approximation for both the ribbon and the gas
molecules. The Hamiltonian of the system can be expressed as
\begin{equation}
\hat{\mathcal{H}}=\hat\mathcal{{H}}_R+\hat\mathcal{{H}}_M+\hat{V}\
,
\end{equation}
where $\hat{\mathcal{H}}_R$ corresponds to the pristine 10-AGNR
(before gas adsorption) and can be written as
\begin{equation}
\hat\mathcal{{H}}_R=\sum_{i}\tilde{\epsilon}_{i}\hat{c}_{i}^\dag\hat{c}_{i}
-\sum_{\langle{i},{j}\rangle}\tilde{t}_{{i}{j}}\hat{c}_{i}^\dag\hat{c}_{j}
\ ,
\end{equation}
where the operator $\hat{c}_{i}^\dag$ ($\hat{c}_{i}$) creates
(annihilates) an electron at site $i$ of the ribbon,
$\tilde{\epsilon}_{i}$ is the on-site energy and
$\tilde{t}_{{i}{j}}$ is the hopping integral. Since there is no
spin effect, we have suppressed the spin indices here.

Because of the hydrogen passivation of the edge carbon atoms, both
$\tilde{\epsilon}_{i}$ and $\tilde{t}_{{i}{j}}$ of the edges can
change. Here, we only consider the variation in the hopping
integrals and we set $\tilde{\epsilon}_{i}=0$ as an origin of the
energy. It is reported that the hopping integral between
$\pi$-orbitals at the edges of AGNRs increases about
$\delta\tilde{t}_{{i}{j}}=12\,\%\,t_R$ \cite{Son1}. Therefore, we
set $\tilde{t}_{{i}{j}}=1.12\,t_R$ for the edges and
$\tilde{t}_{{i}{j}}=t_R$ with $t_R=2.66$ eV for the interior
carbon atoms \cite{Chico}. Also, the Hamiltonian of the adsorbed
molecules, $\hat\mathcal{H}_M$, is obtained by summation over the
Hamiltonian of the isolated molecules ($\hat h_\ell$ for molecule
$\ell$) as:
\begin{equation}\label{V}
\hat\mathcal{{H}}_M=\sum_{\ell}\hat h_\ell \ ,
\end{equation}
with
\begin{equation}
\hat{h}_\ell=\sum_{n_\ell}\epsilon_{n_\ell}\hat{d}_{n_\ell}^\dag\hat{d}_{n_\ell}
-\sum_{\langle{n_\ell},{m_\ell}\rangle}t_{{n_\ell}{m_\ell}}\hat{d}_{n_\ell}^\dag\hat{d}_{m_\ell}
\ ,
\end{equation}
where $\hat{d}_{n_\ell}^\dag$ ($\hat{d}_{n_\ell}$) creates
(annihilates) an electron on atomic orbital $n$ of the molecule
$\ell$ and for simplicity, the atomic basis orbitals are taken to
be orthonormal. Extension to the non-orthogonal bases is
straightforward in principle. The quantities $\epsilon_{n_\ell}$
and $t_{{n_\ell}{m_\ell}}$ are the on-site energy and hopping
integral parameters for the atoms in the gas molecule. These
parameters depend on the type and the geometrical arrangements of
atoms in the molecule. For instance, CO$_2$ molecule consists of
two type atoms, carbon and oxygen, and there is no bonding between
two oxygen atoms in the molecule, i.e. $t_{\mathrm{O-O}}=0$.
Hence, we may have different values for the $\epsilon_{n_\ell}$
and $t_{n_\ell m_\ell}$. Finally, the interaction between the
pristine 10-AGNR and the gas molecules is written as a sum over
each molecule:
\begin{equation}\label{V}
\hat V=\sum_{\ell}\hat v_\ell \ ,
\end{equation}
with
\begin{equation}\label{V}
\hat v_\ell=-\sum_{i=1}^{N_s}\sum_{n=1}^{N_m}t_{i
n}(\hat{c}_{i}^\dag\hat{d}_{n}^{}+\hat{d}_{n}^\dag\hat{c}_{i}
)\delta_{i,i_\ell}\delta_{n,n_\ell}\ ,
\end{equation}
where the indice $n_\ell$ labels the atomic site $n$ of the
molecule $\ell$ that is adsorbed on site $i=i_\ell$ of the ribbon,
and $t_{in}=\tau$ is the hopping integral between the ribbon and
the molecule. $N_m$ is the number of atoms at each molecule.

\begin{figure}
\centerline{\includegraphics[width=1.0\linewidth]{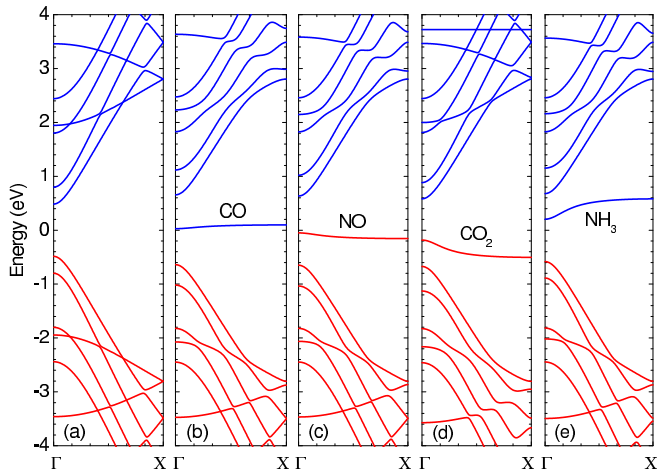}}
\caption{(Color online) Electronic band structure of 10-AGNRs (a)
before and (b)-(e) after gas molecule adsorption: (b) CO, (c) NO,
(d) CO$_2$, (e) NH$_3$. In the case of adsorption, each unit cell
contains a single molecule connected to one of the ribbon edges.
$\Gamma$ and X points correspond to the wave numbers $k=0$ and
$k=\pm\pi/\sqrt{3}a$ in the first Brillouin zone of the ribbon,
respectively.}
\end{figure}

Now we can calculate the Green's function of the whole system to
obtain the average LDOS. For this purpose, as mentioned above, we
divide the system into an infinite series of supercells. The
supercells are labeled by $\ell=0,\pm 1, \pm 2, \cdots$ and we
focus on one of them, say $\ell$=0, and consider the supercells
$\ell=-1,-2,\cdots$ and $\ell=+1,+2,\cdots$ as neighboring
supercells at the left and right, respectively. Based on the
iterative procedure introduced by L\'{o}pez Sancho \textit{et al}.
\cite{Lopez}, the Green's function of each supercell (in the
presence of the gas molecules), due to the periodicity of the
system, can be expressed as
\begin{equation}\label{G00}
\hat{G}(E)=(\hat{\epsilon}-\hat{H}_{00}-\hat{H}_{01}\hat{T}-\hat{H}^\dagger_{01}\hat{\bar{T}})^{-1}\
,
\end{equation}
where $\hat{\epsilon}=(E+i\eta)\hat{I}$ with $\eta$ being a
positive infinitesimal number and $\hat{I}$ is the identity
matrix. $\hat{H}_{00}$ represents the coupling between atoms
within each supercell and $\hat{H}_{01}$ stands for the
interaction between two nearest-neighbor supercells. We note that,
both are matrices of dimension $(N_m+N_s)\times(N_m+N_s)$. If we
designate the Hamiltonian of a single supercell (say, $\ell=0$) of
the pristine 10-AGNR by $\hat\mathcal{{H}}^{(00)}_{R}$, then
$\hat{H}_{00}$ can be written in matrix form as
\begin{equation}
\hat{H}_{00}=\left(\begin{array}{cc}
\hat h_0 & \hat v_0 \\
\hat v^\dagger _0 & \hat\mathcal{{H}}^{(00)}_{R}\\
\end{array}\right)\  ,
\end{equation}
and
\begin{equation}
\hat{H}_{01}=\left(\begin{array}{cc}
\hat{0} & \hat{0} \\
\hat\mathcal{{H}}^{(01)}_R & \hat{0}\\
\end{array}\right)\  ,
\end{equation}
where $\hat{0}$ is a ${(N_m+N_s-10)\times(N_m+N_s-10)}$ zero
matrix and $\hat\mathcal{{H}}^{(01)}_R$ corresponds to the
interaction between nearest-neighbor zigzag lines (perpendicular
to the periodic direction) in the $\ell=0$ and $\ell=1$
supercells. Also, $\hat{T}$ and $\hat{\bar T}$ are the transfer
matrices which can be easily computed using the matrices
$\hat{H}_{00}$ and $\hat{H}_{01}$ through the iterative procedure
\cite{Lopez},
\begin{eqnarray}\label{T1}
\hat{T}&=&P_0+\tilde{P}_0P_1+\tilde{P}_0\tilde{P}_1P_2+\cdots+\tilde{P}_0\cdots
\tilde{P}_{q-1}P_q\nonumber \ ,\\
\hat{\bar{T}}&=&\tilde{P}_0+P_0\tilde{P}_1+P_0P_1\tilde{P}_2+\cdots+P_0\cdots
P_{q-1}\tilde{P}_q\nonumber\ ,
\end{eqnarray}
where $P_k$ and $\tilde{P}_k$ can be determined from the following
recursion relations:
\begin{eqnarray}
P_k&=&(\hat{I}-P_{k-1}\tilde{P}_{k-1}-\tilde P_{k-1}
P_{k-1})^{-1}P_{k-1}^2\nonumber\ ,\\
\tilde{P}_k&=&(\hat{I}-P_{k-1}\tilde{P}_{k-1}-\tilde P_{k-1}
P_{k-1})^{-1}\tilde{P}^2_{k-1}\ ,
\end{eqnarray}
and
\begin{eqnarray}
P_0&=&(\hat{\epsilon}-\hat{H}_{00})^{-1}\hat{H}_{01}^\dag\  ,\nonumber\\
\tilde{P}_0&=&(\hat{\epsilon}-\hat{H}_{00})^{-1}\hat{H}_{01}\  .
\end{eqnarray}
The process can be performed until the desired precision is
achieved, that is, until $P_q,\tilde{P}_q\leq \delta$ with
$\delta$ arbitrarily small.

To study the electronic properties of the system, we obtain the
diagonal matrix elements of the supercell Green's function at the
ribbon sites and average over all configurations. Therefore, the
the average LDOS per carbon atom can be expressed as
\begin{equation}
\langle{D}(E)\rangle=-\frac{1}{\pi
N_s^2}\sum_{i_0=1}^{N_s}\sum_{i=1}^{N_s}\,\mathrm{Im}\,G_{i_0}(i,i;E)\
,
\end{equation}
where $G_{i_0}(i,i;E)$ corresponds to the diagonal matrix elements
of $\hat{G}(E)$ at the ribbon sites, when the gas molecule is
contacted to the ribbon at site $i=i_0$.

\begin{figure}
\centerline{\includegraphics[width=0.95\linewidth]{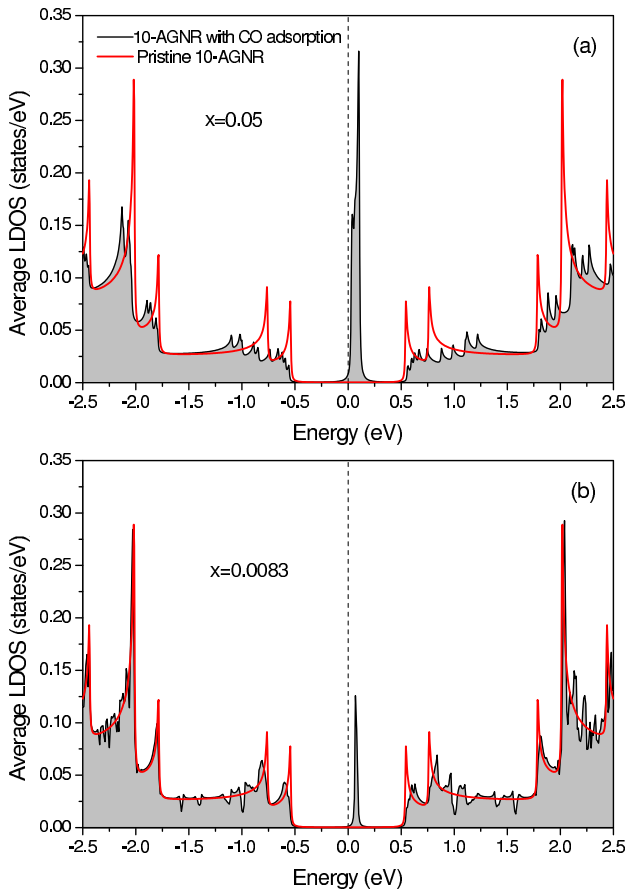}}
\caption{(Color online) Average LDOS of 10-AGNR with CO adsorption
for two values of the gas concentration, in comparison with the
LDOS of pure system. The dashed line shows the original ribbon
Fermi energy.}
\end{figure}

\begin{figure}
\centerline{\includegraphics[width=0.95\linewidth]{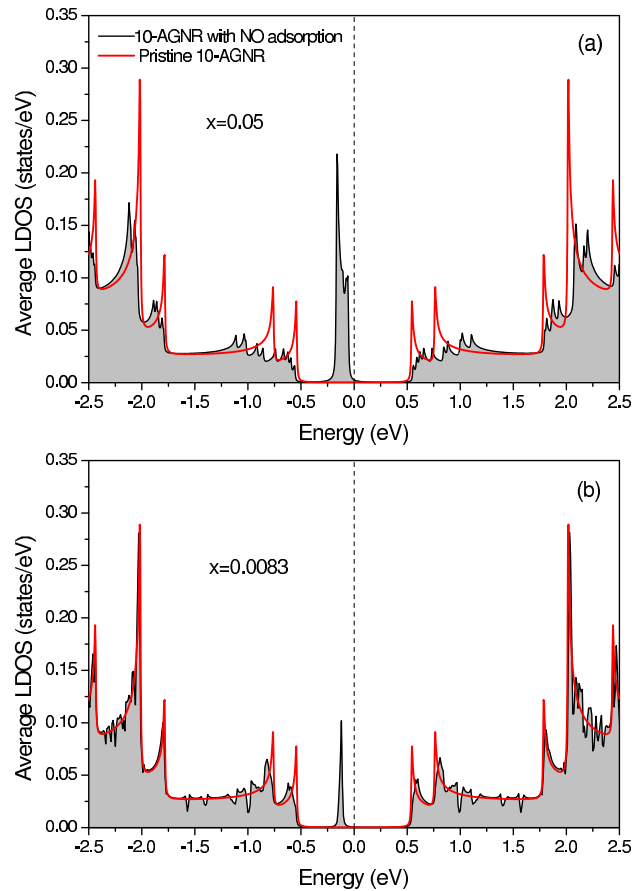}}
\caption{(Color online) The same as Fig. 3 but for NO adsorption.}
\end{figure}

\section{Results and discussion}
We now use the model described above to study gas sensing
properties of the system. The results of numerical computations
will be presented for CO, NO, CO$_2$, and NH$_3$ gas molecules
adsorbed on the 10-AGNR. These simple molecules have been chosen
due to their different electronic properties in the adsorption
process on the ribbon. In order to obtain the hopping parameters
related to the adsorbates, we use of the scaling form of the
tight-binding hopping integrals as \cite{Harrison}
\begin{equation}\label{Har}
t_{\alpha\beta}\simeq t_{R}(\frac{d_{R}}{d_{\alpha\beta}})^2\  ,
\end{equation}
where $d_R=1.42$ {\AA} is the bond length between carbon atoms in
graphene \cite{Saito}, while $d_{\alpha\beta}$ can be the bond
length between the ribbon and the molecule, i.e.
$\alpha\in\mathrm{\emph{ribbon}}$ and
$\beta\in\mathrm{\emph{molecule}}$, or between the atoms in the
molecule, i.e. $\alpha$ and $\beta\in\mathrm{\emph{molecule}}$. We
should note that, this formula is a valid approximation for the
variation of the hopping parameter for a bond between two specific
atoms when the distance between the atoms is slightly varied. In
spite of this, we believe that, by scaling all the hopping
parameters with $t_R$, this formula is able to qualitatively
reproduce the electronic states of adsorbates similar to the
\textit{ab initio} results. To calculate the exact values for
$t_{\alpha\,\beta}$, we need $d_{\alpha\,\beta}$ parameters.
Accordingly, we use the DFT results for these bond lengths to
investigate the adsorption effects of the above-mentioned
molecules \cite{Huang}. Moreover, the exact values for
$\epsilon_{n_\ell}$ are found with varying the on-site energies to
obtain the desired LDOS which is consistent with the reported DFT
results \cite{Huang}. Below we present the values of these
parameters and discuss the adsorption effects of each adsorbate on
the ribbon separately. Note that, we can also use the results of
Ref. \cite{Leenaerts} for comparison, in addition to Ref.
\cite{Huang}, because in this study the gas molecules can be
adsorbed by each carbon atom across the ribbon width, contrary to
Ref. \cite{Huang}.

\subsection{CO on 10-AGNR}
We first emphasize that, according to the DFT calculations
\cite{Leenaerts}, the size of charge transfer between the
molecules and the ribbon depends on the orientation of the
molecules with respect to the ribbon surface. Moreover, the
highest occupied molecular orbital in CO molecule is located on
the C atom. Therefore, CO molecule is attached to a single atom of
the ribbon via its carbon atom with the bond distance
$d_{\mathrm{C-C}}=1.35$ \AA, while the bond length of the adsorbed
CO is $d_{\mathrm{C-O}}=1.18$ \AA\,\cite{Huang}. Inserting these
values into Eq. \ref{Har}, we obtain $\tau=1.11\,t_R$, and
$t_{\mathrm{C-O}}=1.45\,t_R$. Also, the appropriate on-site
energies are given by $\epsilon_\mathrm{C}=-1.40\,t_R$ and
$\epsilon_\mathrm{O}=-1.40\,t_R$. Hence, the effect of this
adsorbed molecule on the electronic structure of the system can be
calculated numerically.

The band structure and the average LDOS of 10-AGNR with CO
adsorption are shown in Figs. 2(b) and 3, respectively. The band
structure, which corresponds to gas concentration $x=0.05$, has
been calculated for a configuration in which the molecule is
connected to one of the ribbon edges. In comparison with Fig. 2(a)
for the pristine 10-AGNR, it induces an impurity sate in the band
gap, slightly above the Fermi energy of the clean system. To see
the effects of all configurations in the presence of finite
concentrations of gas molecule, we have plotted the average LDOS
for two concentrations; $x=0.05$ which indicates that each
supercell with one unit cell contains a single CO molecule, as
shown in Fig. 3(a), and $x=0.0083$ which indicates that each
supercell with six unit cells contains a single CO molecule as
shown in Fig. 3(b).

\subsection{NO on 10-AGNR}
In adsorption of NO molecules on the 10-AGNR, only N atoms are
attached to the ribbon \cite{Huang}, with the bond distance
$d_{\mathrm{C-N}}=1.43$ \AA, while the bond length of the adsorbed
NO is $d_{\mathrm{N-O}}=1.24$ \AA. Accordingly, using Eq.
\ref{Har} the hopping energies are given as $\tau=0.98\,t_R$ and
$t_{\mathrm{N-O}}=1.31\,t_R$. In addition, the appropriate values
for the on-site energies are $\epsilon_\mathrm{N}=-1.42\,t_R$ and
$\epsilon_\mathrm{O}=-1.35\,t_R$. In this case, a single NO
molecule on one of the ribbon edges induces an impurity sate in
the band gap, slightly below the Fermi energy of the pure system
[see Fig. 2(c)]. The results of LDOS clearly indicate that the
values of charge density of the pristine 10-AGNR are modulated by
averaging over all adsorption configurations. This type of
adsorbate, similar to CO molecule, does not change the position of
the bottom (top) of the conduction (valence) band.

From the average LDOS in Figs. 3 and 4, we notice that the
adsorption of both CO and NO molecules by the AGNR decreases the
original band gap, especially at high $x$, and may affect the
optical properties of the system. At low gas concentration, such
as $x=0.0083$, the density of states of gas molecules consists of
a $\delta$-function peak near the center of band gap, while at
high $x$, such as $x=0.05$, a narrow band appears instead of a
level. However, this narrow band is far from the bottom (top) of
the conduction (valence) band and one expects that such a gas
adsorption does not enhance the charge transport of the system.

In the case of CO and NO adsorption, we expect that the Fermi
level does not change considerably due to the position of
molecular level inside the band gap with respect to the edges of
valence and conduction bands. As we know, in a non-degenerate
$n$-type ($p$-type) semiconductor, the Fermi energy is rather far
from the edge of the conduction (valence) band. This feature
indicates that the 10-AGNR with adsorbed molecules like CO and NO
behaves as a non-degenerate semiconductor.
\begin{figure}
\centerline{\includegraphics[width=0.95\linewidth]{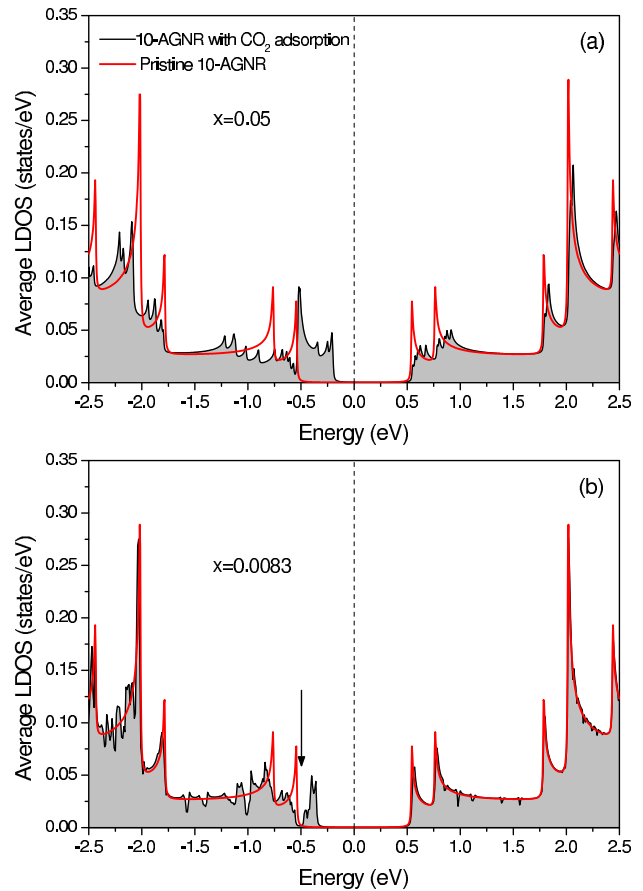}}
\caption{(Color online) The same as Fig. 3 but for CO$_2$
adsorption.}
\end{figure}

\begin{figure}
\centerline{\includegraphics[width=0.95\linewidth]{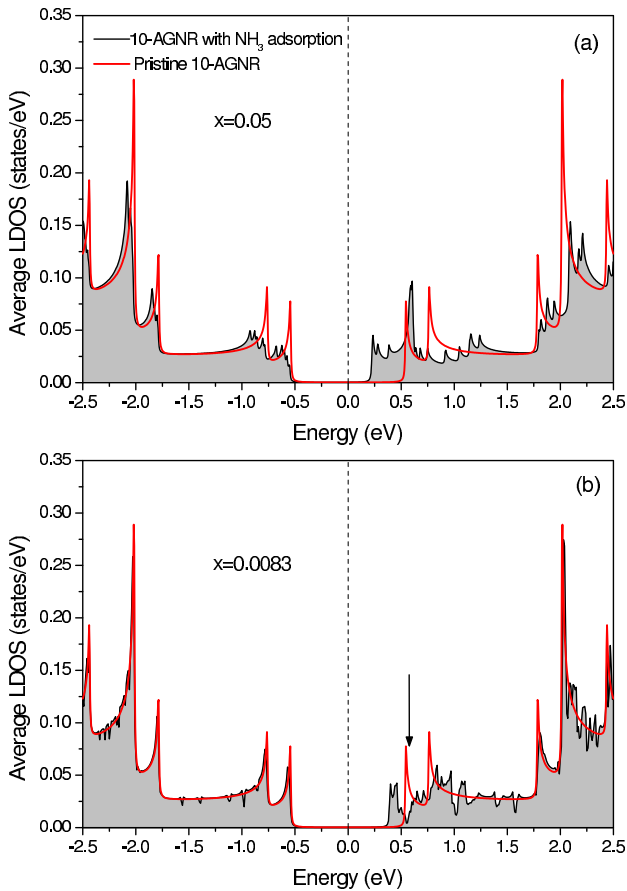}}
\caption{(Color online) The same as Fig. 3 but for NH$_3$
adsorption.}
\end{figure}

\subsection{CO$_2$ on 10-AGNR}
The DFT results of CO$_2$ adsorbed on the ribbon show that the
molecule sits 1.51\,\AA\, away from the edge carbon atoms and the
length of C-O bonds is 1.26\,\AA\, \cite{Huang}. Accordingly, the
calculated hopping energies are given as $\tau=0.89\,t_R$ and
$t_{\mathrm{C-O}}=1.27\,t_R$. Moreover, we choose
$\epsilon_\mathrm{C}=1.75\,t_R$ and
$\epsilon_\mathrm{O}=1.40\,t_R$ as the suitable values for the
on-site energies. Using these parameters, the band structure for
the configuration where the molecule sits on one of the edges is
given by Fig. 2(d). An interesting feature of this figure is that
the molecular states form an impurity band in the original band
gap with width 0.3 eV. We see that, due to the type of charge
exchange between the gas molecule and the ribbon, the energy
levels of the molecule are close to the edge of the valence band.
Also, the average LDOS analysis clearly show that, at $x=0.05$
(Fig. 5(a)) the impurity band overlaps with the valence band and
consequently the band gap reduces significantly. On the other
hand, for low enough doping, such as $x=0.0083$, the impurity
states become more localized on the gas molecules, the adsorbate
bandwidth decreases and a gap opens between impurity and the top
of the valence bands [see the arrow in Fig. 5(b)]. This type of
molecular doping can enhance the charge transport in this system
and hence the CO$_2$ molecule will act as an acceptor at finite
$x$ and temperature.

\subsection{NH$_3$ on 10-AGNR}
The \emph{ab} \emph{initio} studies of gas adsorption on graphene
\cite{Leenaerts} and 10-AGNR \cite{Huang} indicate that in the
ammonia molecule, the N atom is closer to the surface than the H
atoms. The bond lengths of the adsorbed NH$_3$ are given as
$d_{\mathrm{C-N}}=1.49$ \AA\, and $d_{\mathrm{N-H}}=1.03$
\AA\,\cite{Huang}. Therefore, the hopping parameters are $\tau=
0.9\,t_R$ and $t_{\mathrm{N-H}}=1.9\,t_R$, and the on-site
energies are chosen to be $\epsilon_\mathrm{N}=-3.45\,t_R$ and
$\epsilon_\mathrm{H}=-2.65\,t_R$. Fig. 2(e) shows the electronic
band structure of the system where the molecule sits on one of the
ribbon edges. In this case the impurity band is close to the
bottom of the conduction band with width 0.38 eV.

We see in Fig. 6(a) that in the case of $x$=0.05, by averaging
over all gas adsorption configurations, the impurity band overlaps
with the conduction band and the band gap decreases considerably.
In addition, at $x$=0.0083 as shown in Fig. 6(b), the impurity
bandwidth decreases and at low enough $x$ the impurity band is
separated from the conduction band and a gap gradually opens [see
the arrow in Fig. 6(b)]. This indicates that NH$_3$ molecule acts
as a donor, which is in good agreement with experimental
\cite{Schedin} and theoretical studies of NH$_3$ adsorption on
graphene surface \cite{Leenaerts} and 10-AGNR \cite{Huang}.

From the above discussions we conclude that, Co$_2$ or NH$_3$
molecules adsorption shifts the Fermi energy towards and even into
the original valence or conduction band, and the system exhibits
$p$-type or $n$-type semiconducting behavior. On the other hand,
when the Fermi energy is close to or even higher (lower) than the
bottom (top) of the conduction (valence) band, we have a so called
degenerate semiconductor. Therefore, by looking at the Figs. 5 and
6, one can conclude that the adsorption of CO$_2$ or NH$_3$
molecules converts the electronic structure of 10-AGNR as a
degenerate semiconductor. We should note that, this feature which
strongly depends on the gas concentration has not been
investigated in the reported DFT calculations.

By comparing Figs. 2(b)-2(e) which correspond to the gas
adsorption configuration on one of the ribbon edges with Figs. 3-6
which correspond to the average of LDOS over all configurations,
we conclude that the interaction between gas molecules and the
carbon atoms at the ribbon edges is much stronger than that of the
interior carbon atoms, which is consistent with the DFT results
\cite{Huang}.
\section{Conclusion}

In this research, using the parameters obtained from DFT results,
we have modeled the adsorption of gas molecules CO, NO, CO$_2$,
and NH$_3$ on the 10-AGNR. Under the assumption that the variation
of bond lengths between carbon atoms is only localized on the
edges of ribbon, and including this effect in our calculations, we
studied the influence of finite concentrations of the adsorbates
on the average LDOS. In addition, we assumed that only one
molecule can occupy the ribbon sites and there is no interaction
among the adsorbed molecules.

The results show that, although all types of gas molecules can
influence the electronic properties of the system, the average
LDOS varies from one type to other. In the cases of CO and NO
molecules adsorption, the impurity states are very localized at
the center of original band gap. Therefore, adsorption of these
gas molecules on 10-AGNR cannot notably affect the charge
transport through such systems. In the cases of CO$_2$ and NH$_3$
adsorption, however, the system exhibits $p$-type and $n$-type
semiconducting behavior, respectively. Despite the simplicity of
the model, reasonable agreement with DFT results is demonstrated.
The results can be useful in designing future gas sensors and
$p$-type or $n$-type semiconductors.

\section*{Acknowledgement}
The author thanks F.M. Peeters for valuable comments and M. Farjam
for useful discussions. This work was supported by Payame Noor
University grant.

\end{document}